\documentclass{iau} 
\usepackage{graphicx}

\title[FIR spectroscopy of the Galactic Center] 
{FIR Spectroscopy of the Galactic Center: Hot and Warm Molecular Gas}

\author[Goicoechea et al.]   
{Javier R. Goicoechea$^1$, Mireya Etxaluze$^{1}$, Jos\'e Cernicharo$^{1}$, 
Maryvonne Gerin$^{3}$, Jerome Pety$^{4}$ and collaborators
}

\affiliation{$^1$ICMM-CSIC, Cantoblanco, Madrid.
E-28049, Spain  \\[\affilskip]
$^3$LERMA, Observatoire de Paris, CNRS UMR 8112, \'Ecole Normale Sup\'erieure, France.\\[\affilskip]
$^4$IRAM, F-38406, Saint Martin d'H\`eres, France.\\
}

\pubyear{2016}
\volume{322}  
\jname{The Multi-Messenger Astrophysics of the Galactic Centre}
\editors{Steve Longmore, Geoff Bicknell and Roland Crocker, eds.}
\begin{document}

\maketitle

\begin{abstract}

The angular resolution ($\sim$10$''$) 
achieved by the Herschel Space Observatory $\sim$3.5\,m telescope  at FIR wavelengths allowed us to roughly separate the emission toward the inner  parsec of the  galaxy (the central cavity) from that of the surrounding circumnuclear disk (the CND). 
The FIR spectrum toward Sgr~A$^*$ is dominated by intense [O\,{\sc iii}], [O\,{\sc i}], [C\,{\sc ii}], [N\,{\sc iii}], [N\,{\sc ii}], and [C\,{\sc i}] 
fine-structure lines (in decreasing order of luminosity) arising in  gas irradiated
 by the strong UV field from the central stellar cluster. The
 high-$J$ CO rotational line intensities observed at the interface between the inner CND 
 and the central cavity are consistent with a hot isothermal component 
at $T_{\rm k}$$\approx$10$^{3.1}$\,K and $n$(H$_2$)$\approx$10$^4$\,cm$^{-3}$. 
They are also consistent with a distribution of lower temperatures at higher 
gas density, with most CO at $T_{\rm k}$$\approx$300\,K. 
The hot CO component (either the bulk of the CO column density or just a small fraction 
depending on the above scenario) likely results from a combination of UV and 
shock-driven heating. If \mbox{UV-irradiated} and heated dense clumps do not exist,  shocks likely dominate the heating  of the hot molecular gas component. 
Although this component is beam diluted in our FIR observations, it may be resolved
at much higher angular resolution.
An ALMA project using different molecular tracers to characterize  UV-irradiated shocks in 
the innermost layers of the CND is ongoing.

\end{abstract}

\firstsection 

\section{The neutral gas component of the Galactic Center}

The interstellar material within a few parsecs from the
central supermassive black hole of the Milky Way (near the Sgr\,A$^*$~radio source position) is a unique template for our understanding of galactic nuclei and galaxy evolution.
Widespread shocks, high-energy radiation, enhanced magnetic fields and strong tidal forces, all shape a very singular  environment.
The distribution of interstellar gas and dust around Sgr\,A$^*$ 
consists of a central cavity of $\sim$1\,pc  radius containing
 warm dust and ionized gas.
Some of the ionized gas streamers (the ``mini-spiral'') bring material
close to the very center.
Between $\sim$1.5\,pc and $\sim$5\,pc, a disk of dense molecular gas exists
(the CND).
However, its physical and kinematical properties
are not fully constrained, thus it is not yet clear whether all the material
in the  CND is stable against the strong tidal forces 
in the region, or has a more transient nature 
(e.g., \cite[Requena-Torres et al. 2012]{Requena12} and references therein).

The interface between the inner CND and the central cavity likely contains a component
of warm neutral gas,  detectable through FIR atomic fine structure line emission and   through molecular line emission/absorption features. 
The nature of this component, its ionization sources, heating mechanisms  
 and kinematic patterns (in-falling gas from the CND? outflows? orbiting material?) are not known.

\begin{figure}[t]
\begin{center}
 \includegraphics[width=5.25in]{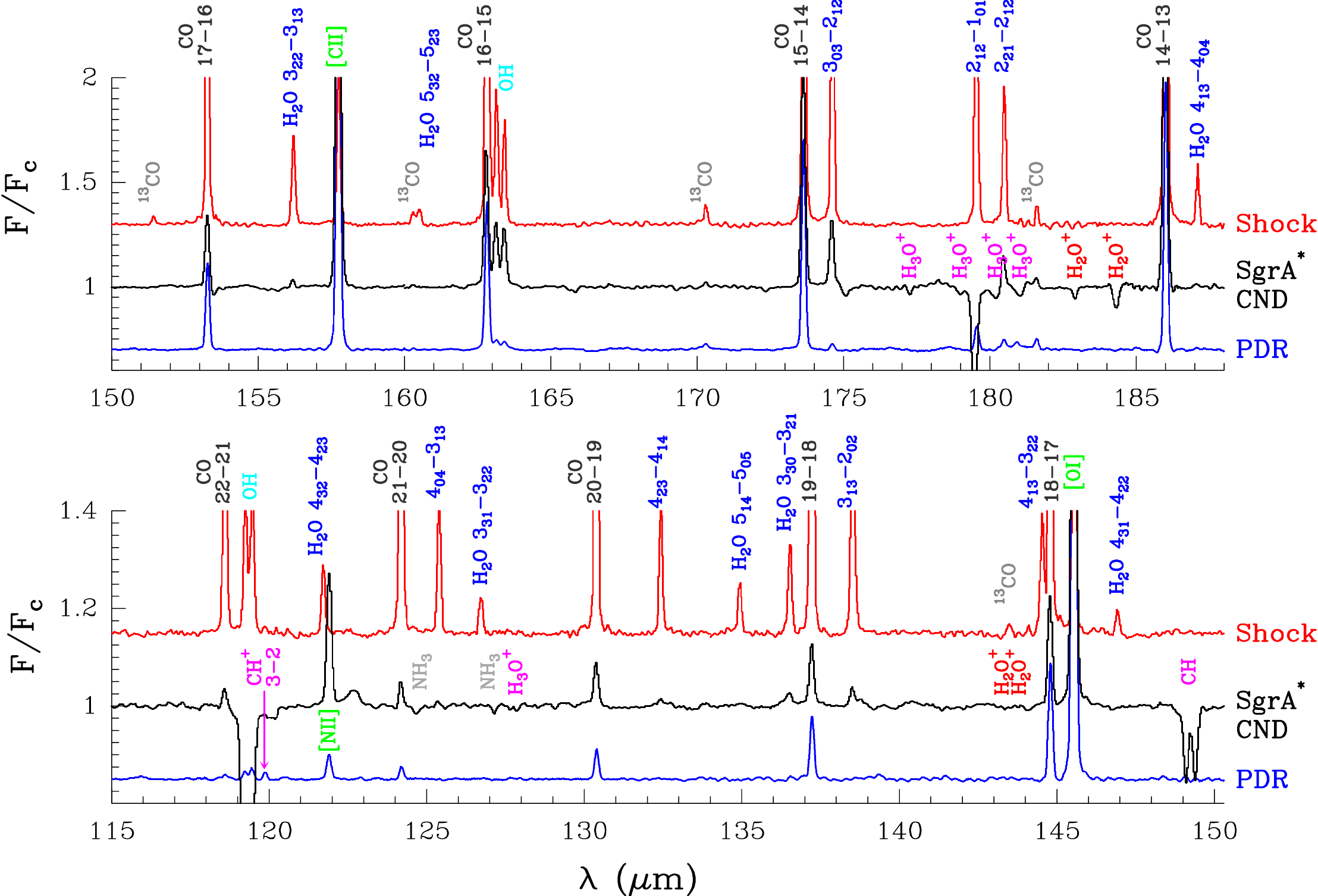} 
 \caption{Comparative FIR spectra of three different template environments in the Milky Way taken with
\textit{Herschel}/PACS. They correspond, from top to bottom, to 
\mbox{Orion~BN/KL outflows} (red; \cite[Goicoechea et al. 2015]{Goicoechea15}), 
the CND around Sgr\,A$^*$ (black;  
\cite[Goicoechea et al. 2013]{Goicoechea13}), 
and the Orion Bar PDR (blue). The ordinate scale refers to the line flux to continuum flux ratio.}
\label{fig1}
\end{center}
\end{figure}

Owing to the small extinction effects at FIR wavelengths, 
and because of the strong emission from the ISM component related to 
AGN and star formation, the relevance of FIR spectroscopy ($\sim$50-350\,$\mu$m) to characterize extragalactic nuclei has notably increased thanks to Herschel, and now to ALMA  observations of high-$z$ galaxies.
\cite[Goicoechea et al. (2013)]{Goicoechea13} presented the complete FIR/submm spectrum toward  Sgr\,A$^*$. 
The emission is dominated by  strong 
[O{\sc iii}],  [O{\sc i}],  [C{\sc ii}], [N{\sc iii}],  [N{\sc ii}],  and [C{\sc i}]  fine structure  lines.  In addition, rotationally excited lines of CO
(from $J$=4-3 to 24-23), H$_2$O, OH, H$_3$O$^+$, HCN and HCO$^+$ (up to $J$=8-7) 
as well as  ground-state absorption lines of OH$^+$, H$_2$O$^+$, 
CH$^+$ and HF are detected.  Figure~1 shows a comparison of FIR 
spectra of three archetypal environments in the Milky Way: a protostellar outflow and associated shocked molecular gas,  a sight line to the CND including dense gas and foreground absorption  by diffuse gas, and a highly UV-irradiated dense PDR. 
Whereas the shocked region  shows very intense rotationally excited CO, H$_2$O and OH  
emission lines (in decreasing order of luminosity), the PDR spectrum is dominated by strong [C\,{\sc ii}] and [O\,{\sc i}] emission, with faint lines from H$_2$O. The warm neutral gas in the Galactic
 Center has a richer spectrum, with hydrides such as NH$_3$, OH, H$_2$O,  H$_3$O$^+$, and  H$_2$O$^+$ indicating 
the presence of UV- and CR-irradiated  gas. These molecules are powerful tracers of the excitation conditions
 and ionization sources of the  warm molecular gas
(see review by \cite[Gerin et al. 2016]{Gerin16}).

\end{document}